\documentclass[prd,reprint,a4paper,showpacs,superscriptaddress,nofootinbib,floatfix]{revtex4-1}

\newcommand{\beq}{\begin{equation}}
\newcommand{\eeq}{\end{equation}}
\newcommand{\beqa}{\begin{eqnarray}}
\newcommand{\eeqa}{\end{eqnarray}}

\def\bc{\begin{center}}
\def\ec{\end{center}}

\def\gaga{\gamma\gamma}

\usepackage{graphics}
\usepackage{graphicx}
\usepackage{amssymb}
 \usepackage{amsthm}

\begin{document}
\title{Neutralino dark matter and Higgs mediated lepton flavor violation
in the minimal supersymmetric standard model
}

\author{M.~Cannoni}
\affiliation{Universit\`a di Perugia, Dipartimeno di Fisica,
Via A.~Pascoli, 06123, Perugia, Italy}

\author{O.~Panella}
\affiliation{Istituto Nazionale di Fisica Nucleare, Sezione di Perugia, Via A.~Pascoli, 06123 Perugia, Italy}

\date{\today}

\begin{abstract}
We re-examine the prospects for the detection of Higgs mediated lepton flavor violation at
LHC, at a photon collider and in $\tau$ decays 
such as $\tau\to\mu\eta$, $\tau\to\mu\gamma$.
We allow for the
presence of a large, model independent, source of lepton flavor
violation in the slepton mass matrix in the $\tau-\mu$ sector by the
mass insertion approximation 
and constrain the parameter space using the $\tau$ LFV decays
together with the $B$-mesons physics
observables, the anomalous magnetic moment of the muon and the dark matter
relic density. We further impose the exclusion limit on spin-independent
neutralino-nucleon scattering 
from CDMS and the CDF limits from direct search of the heavy neutral Higgs at 
the TEVATRON. 
We find rates
probably too small to be observed at future experiments if models have to 
accommodate for 
the relic density measured by WMAP and explain the $(g-2)_{\mu}$ anomaly:
better prospects are found if these two constraints are applied
only as upper bounds.  
The spin-independent
neutralino-nucleon cross section in the studied constrained parameter space
is just below the present CDMS limit and the running XENON100
experiment will cover the region of the parameter space where the
lightest neutralino has large gaugino-higgsino mixing.

\end{abstract}

\maketitle

\section{Introduction}
\label{intro}

One of the appealing features of the minimal supersymmetric standard model (MSSM)
with R-parity conservation is the presence of a neutral, stable particle, 
the lightest neutralino, which presents all the characteristics to be a possible
candidate for accounting for the cold dark matter in the Universe
~\cite{kamionkowski}. 
The amount of dark matter $\Omega h^2$, where $\Omega$ is the dark matter density normalized
to the critical density of the Universe and $h$ is the reduced 
Hubble constant, recently has been precisely measured by the WMAP experiment~\cite{wmap}. 

The Higgs sector of the MSSM~\cite{djouadi}, especially the 
heavy neutral Higgses $A$ and $H$, play a prominent role in the physics
of neutralino dark matter in two ways.
In some region of the supersymmetric (SUSY) parameter space
neutralinos yield the desired amount of relic density by annihilating 
into fermions through the $s$-channel resonant exchange of neutral Higgs bosons $h$, $H$, $A$,
the so called funnel region where $m_A \simeq 2 m_{\chi}$. 
As dark matter is expected to be distributed 
in a halo surrounding our galaxy,  
neutralinos can scatter off nuclei in terrestrial detectors: coherent scattering is mediated
by scalar interactions through the $s$-channel exchange of squarks and 
$t$-channel exchange of the CP-even neutral 
Higgs bosons $h$ and $H$. 
These effects become sizeable when squarks are heavy 
and $\tan\beta$ is large in reason of the enhanced Higgs bosons coupling to
down-type fermions, especially for the $b$ quark which has the
largest Yukawa coupling: moreover this couplings receive 
large radiative SUSY-QCD corrections at large $\tan\beta$
that can be relevant for their production in hadron-hadron
collisions at TEVATRON and LHC~\cite{carena1,carena2,carena3}.
In this scenario it is well known that 
$B$-mesons physics observables are very sensible to Higgs physics 
~\cite{bkq,isidori1,isidori2,buras,isidori3} and put strong constraints
on the parameter space.
The branching ratios for $\tau$ lepton flavor violating decays
are also enhanced near the experimental bounds
~\cite{bkl,sher,dedes,brignole1,brignole2,arganda,kanemura2,parry,paradisi2,paradisi1,chuan,arganda2,herrero}.

Once a source of LFV is present in the slepton mass matrix, for example
the MSSM with the celebrated see saw mechanism for generation of small 
neutrino masses,
two different mechanisms of LFV arise:
gauge-mediated LFV effects through the exchange of gauginos and sleptons
~\cite{borzumati,hisano1,hisano2}
and Higgs-mediated LFV effects through effective non-holomorphic
Yukawa interactions for quarks and leptons
~\cite{bkq,bkl}.
LFV Yukawa couplings of the type $\bar{L}_R^i L_L^j H_u^*$ are induced at loop
level and become particularly sizable at large $\tan\beta$.
In this case the effective flavor-violating Yukawa interactions are described by the
lagrangian:
\beqa
\mathcal{-L}&\simeq&(2G_{F}^2)^{1/4}\frac{m_{l_i}} {\cos^2{\beta}}
\left(\Delta^{ij}_{L} {\overline{l}}^i_R  l^j_L + \Delta^{ij}_{R}
{\overline{l}}^i_L  l^j_R \right)\cr
&\times&\left(\cos({\beta-\alpha})h-\sin({\beta-\alpha})H-iA \right) +h.c.
\label{lagrangian}
\eeqa
where $\alpha$ is the mixing angle between the CP-even Higgs bosons $h$ and
$H$, $A$ is the physical CP-odd boson,
$i,j$ are flavor indices that in the following are understood to be different ($i\neq j$).
The coefficients $\Delta^{ij}$ in Eq.~(\ref{lagrangian}) are induced at one loop level
by the exchange of gauginos and sleptons, provided a source of slepton mixing is present.
The expressions
of $\Delta^{ij}_{L,R}$ in the mass insertion approximation are
given by~\cite{paradisi1}:
\beqa
\Delta^{ij}_{L} =
&-&\frac{{g'}^2}{16\pi^2}\,\mu \, m_1 \, \delta^{ij}_{LL} \, m_{L}^2\nonumber\cr
&\times&\left[
I{'} (m_1^2, m_{R}^2, m_{L}^2)+\frac{1}{2} I{'} (m_1^2, \mu^2, m_{L}^2)
\right]\nonumber\cr
&+& \frac{3}{2} \frac{g^2}{16\pi^2} \, \mu \, m_2 \, \delta^{ij}_{LL} \, m_{L}^2
I{'} (m_2^2, \mu^2, m_{L}^2),
\label{deltal}
\eeqa
\beq
\label{deltar}
\Delta^{ij}_{R}=
\frac{{g'}^2}{16\pi^2}\mu \, m_1 \, m^{2}_{R} \, \delta^{ij}_{RR}
\left[I{'}\!(m^{2}_{1},\mu^2,m^{2}_{R})\!-\!(\mu\!\leftrightarrow\! m_{L})
\right]
\eeq
where $g$ and $g'$ are the $SU(2)_L$ and $U(1)_Y$ couplings respectively,
$\mu$ is the the Higgs mixing parameter, 
$m_{1,2}$ the gaugino mass parameters and $m^{2}_{L(R)}$ stands for the left-left
(right-right) slepton mass matrix entry. 
$I'$ is the derivative $dI (x,y,z)/d z$
of the three point one-loop integral.
The LFV mass insertions 
$\delta^{ij}_{LL}\!=\!{({m}^2_{L})^{ij}}/{m^{2}_{L}}$,
$\delta^{ij}_{RR}\!=\!{({m}^2_{R})^{ij}}/{m^{2}_{R}}$,
where $({m}^2_{L,R})^{ij}$ are the off-diagonal flavor changing entries
of the slepton mass matrix, are free parameters which allow for
a model independent study of LFV signals. 


The connection between gaugino-mediated LFV signals and neutralino dark matter 
in the see-saw mechanism implemented in mSUGRA constrained MSSM has been recently studied
in Refs.~\cite{barger1,barger2}: here it is shown that large neutrino Yukawa 
coupling affects the renormalization group evolution equations of SUSY 
parameters from the grand unification (GUT) scale to the electroweak scale
(in a SO(10) GUT scenario) enhancing some LFV rates by orders of magnitude
and changing also the neutralino relic density and direct and indirect 
detection rates.

In this paper, on the other hand, we follow a different phenomenological approach:
the study is done in the framework of MSSM with real parameters assigning the value of 
the parameters at the weak scale without any assumption on the mechanism of SUSY breaking or
the high energy theory, nor on the origin of LFV entries in the slepton mass matrix
and limitate our attention to Higgs 
mediated flavor violating effects. 
We study the interplay between the assumptions of  
the lightest neutralino as dark matter candidate and Higgs 
mediated flavor violation both to constrain the MSSM parameter space
and to give prediction for the neutralino-nucleus
scattering and LFV signals at colliders and in $\tau$ decays. 
For related studies see~\cite{carena3,isidori4,carena4}.

In Section~\ref{sec1} we discuss the 
scan of the parameter space in the real MSSM 
and the imposed constraints.
Than we study their effects 
on the spin-independent
neutralino-nucleon cross section and the arising correlations between
supersymmetric parameters in Sec.~\ref{sec2}.
In Section~\ref{sec3} we analyse LFV signals  $\tau\to\mu\eta$,
$\tau\to\mu\gamma$, $pp\to \Phi \to \tau\mu +X$ at LHC 
and $\gaga \to \tau\mu b \bar{b}$ at a future photon collider.
Conclusions are given in Sec.~\ref{sec4}.   
 
\section{Constrained parameter space }
\label{sec1}

We introduce LFV in the model through the  
mass insertions $\delta_{LL,RR}^{32} =0.5$.
This value ensures the largest rates in LFV
processes and allow us to study the more optimistic scenarios; 
higher values 
contradict the mass insertion approximation as an expansion of propagators in
these small parameters.
Higgs mediated effects become eneteresting at large $\mu$ and $\tan\beta$ and low $m_A$;
further, if SUSY-QCD particles are heavy Higgs effects are dominant also for neutralino 
dark matter.
We thus scan 
the following real MSSM parameter space:
\begin{itemize}
\item[-]
100 GeV $\le m_A \le $1 TeV;
\item[-]
20$\le\tan\beta\le$60;
\item[-]  
500 GeV$\le \mu \le$5 TeV;\\
The sign of $\mu$ is taken positive, as preferred by 
the SUSY explanation of the $(g-2)_\mu $ anomaly. 
\item[-]
150 GeV$\le m_1, m_2\le $1.5 TeV;\\
We do not impose any relation
but let them vary independently. 
To have large masses for gluinos we choose:\\
1 TeV $\le m_3\le$ 5 TeV.
\item[-]
1 TeV$\le m_{U_3}, m_{D_3}, m_{Q_3} \le$5 TeV;\\
for the third generation of squarks: these are 
are varied freely without imposing any relation.
For the first and the second generation the soft masses
are set to be equal, $ m_{U_i} = m_{D_i}
= m_{Q_i} =m_{\tilde{q}}$,
where $i=1,2$ and $m_{\tilde{q}}$ is another free parameter
which varies in the same range. 
\item[-]
300 GeV$\le m_{L_3}, m_{E_3} \le $ 2.5 TeV;\\
for sleptons of the third generation which 
are independent parameters.
For the first and the second generation the soft masses
are set to be equal, $ m_{L_i} = m_{E_i}
=m_{\tilde{\ell}}$, 
where $i=1,2$ and $m_{\tilde{\ell}}$ is another free parameter
which varies in the same range. Sleptons, diffrently from squarks,
can be light in order to explain the $(g-2)_\mu $ anomaly.  
\item[-]
-2$\le \frac{A_{U_3}}{m_{U_3}}$, $\frac{A_{D_3}}{m_{D_3}}$, $\frac{A_{E_3}}{m_{E_3}} \le 2$;\\
for first and second generation the trilinear scalar
couplings are set to zero. 
\end{itemize}

The outlined (16-dimensional) large parameter space is restricted imposing
the following experimental limits:
\begin{figure*}[t!]
\begin{center}
\includegraphics*[scale=0.6]{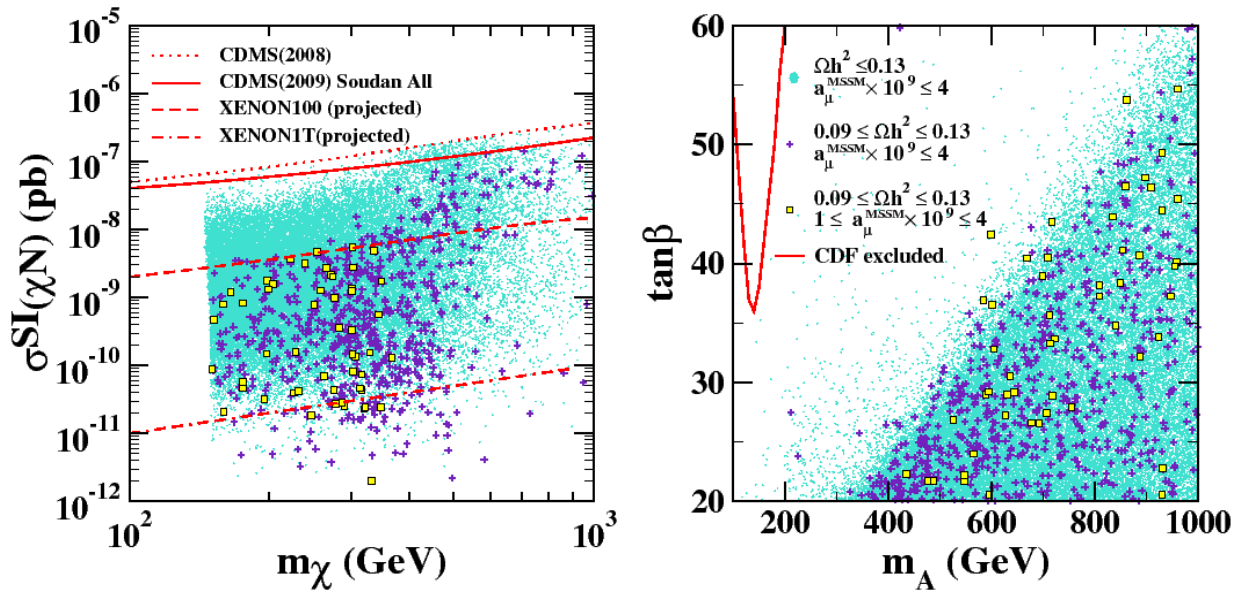}
\caption{
{\it Left:} Scatter plot for the spin-independent neutralino-nucleon
cross section versus the neutralino mass. The area above the solid line is excluded 
the CDMS final results; the area above the dotted line is excluded by the 2008 CDMS
search.
The dashed and dot-dashed lines give the sensitivity reach 
of two phases of the XENON experiment. The scanned parameter space and the imposed experimental 
constraints are described in Sec.~\ref{sec1}. The different graphical presentation
of points corresponds to different steps in imposing the constraints on the 
$(g-2)_\mu /2$ and on the relic density as explained in the legend of right panel.  
{\it Right:} Scatter plot in the ($m_A$, $\tan\beta$) plane. 
The region delimited by the line is excluded by the CDF experiment.}
\label{fig1a}
\end{center}
\end{figure*}

\begin{itemize}
\item[-]{\it Light Higgs and SUSY masses.}\\
LEP, TEVATRON bounds on sparticle masses and
the LEP bound on light Higgs: $m_h \ge$ 114.4 GeV~\cite{pdg}.
\item[-]{\it $B$-physics observables.}\\
For the $B$-physics observables we use the theoretical and experimental numbers
of Table 1 in Ref.~\cite{fit}. Thus we require  
$ 0.995\le \frac{{\cal B}^{MSSM}(B\to X_s \gamma)}{{\cal B}^{SM}(B\to X_s \gamma)}\le 1.239$,
the upper bound on rare decays $B_s \to\mu^+ \mu^-$ branching ratio is set to
$\mathcal{B}(B_s \to\mu^+ \mu^-)\le 4.7\times 10^{-8}$ and
$0.60\le \frac{\Delta m_{B_s}^{MSSM}}{\Delta m_{B_s}^{SM}} \le 1.24$. 
Finally we require
$0.85 \le \frac{{\cal B}^{MSSM}(B\to\tau\nu)}{{\cal B}^{SM}(B\to\tau\nu)} \le 1.65$.
\footnote{There is at present a discrepancy between the SM value and the experimental value
of the purely leptonic decays branching ratio ${\cal B}(B\to\tau\nu)$ due to
a recent analysis~\cite{utfit}.
Given the unclear situation both on the theoretical and experimental side we do not consider it here.}
\item[-]{\it LFV $\tau$ decays.}\\
Once Higgs mediated LFV effects are present in the model
the non-observation of these rare decays puts strong 
constraints on the parameter space. 
The present experimental upper bounds are:
$\mathcal{B}(\tau\to\mu\gamma)\le 4.4\times 10^{-8}$~\cite{babar},
$\mathcal{B}(\tau\to\mu\eta)\le 5\times 10^{-8}$~\cite{belle},
$\mathcal{B}(\tau\to\mu\mu\mu)\le 3.2\times 10^{-8}$~\cite{belle}.
\item[-]{\it Relic density.}\\
We use the conservative WMAP $3\sigma$ interval $0.09\le\Omega h^2\le 0.13$~\cite{wmap}
on the relic density, both
applying only the upper limit, (allowing for other 
sources of dark matter besides the neutralino) and the complete interval.
See the legend of Figure~\ref{fig1a}.
\item[-]{\it Muon anomalous magnetic moment.}\\
The present discrepancy between   
$a_{\mu}^{SM} =(g-2)^{SM}/2$ and the experimental measured value,
$\Delta a_{\mu} = a^{exp}-a_{\mu}^{SM}$,
lies in the interval $(2-4)\times 10^{-9}$~\cite{hagiwara}. 
We always require $a_{\mu}^{MSSM} =(g-2)^{MSSM} /2 \le 4 \times 10^{-9}$. 
We further show the models which satisfy also 
the conservative limit lower bound $a_{\mu}^{MSSM} \ge 1 \times 10^{-9}$. 
See the legend of Figure~\ref{fig1a}.
\item[-]{\it Direct dark matter detection.}\\
The most stringent limit up to date in the neutralino mass range $100-1000$ GeV
for the neutralino-nucleus spin independent cross section
comes from the CDMS experiment. 
The upper limits from the 2008 analysis~\cite{cdms} and 
the recent final combined results~\cite{cdmsfinal}
are  
reported in Fig.~\ref{fig1a} (left panel).
\item[-]{\it Non-standard Higgs  search at TEVATRON.}\\
Recently CDF collaboration has published the most stringent 
exclusion limits in the ($m_A$, $\tan\beta$) plane in the light of
negative results in the  
search for heavy neutral Higgs bosons 
in the inclusive $A,H$ production and the 
successive decay into $\tau^+ \tau^-$ pairs~\cite{cdf}.
The excluded region is limited in the low $m_A$, high $\tan\beta$ region 
and is depicted in Fig.~\ref{fig1a} (right panel).
\end {itemize}
For numerical computations we use 
the code \textsc{DarkSusy}~\cite{darksusy}
for accelerator bounds, the neutralino relic density 
and direct
dark matter detection in our general MSSM. 
\textsc{DarkSusy} uses the code \textsc{FeynHiggs}~\cite{feynhiggs}
for SUSY and Higgs mass spectrum and 
Higgs widths and branching ratios.
For $b\to s\gamma$ we used the routines in \textsc{DarkSusy}
while for 
MSSM contribution to the muon anomalous magnetic moments $(g-2)_\mu$
those of \textsc{FeynHiggs} which include also the leading
and sub-leading two-loop contributions. 
For the others $B$-physics observales we used the formulas of 
Refs.~\cite{buras,isidori3}.
We generate $10^6$ random models, selecting the ones 
which evade the listed constraints. All of them are applied at 
the same time with the exception of  the relic density and $(g-2)_\mu$ anomaly 
for which we also relax the lower bounds: thus 
requiring only $\Omega_{\chi} h^2 \le 0.13$
and $a^{MSSM}_{\mu} \le 4\times 10^{-9}$ 
around $4\times 10^4$ survive, the light grey (turquoise) points in the Figures.
Requiring also $\Omega_{\chi} h^2 \ge 0.09$  around $7\times 10^2$
are left, the plus-shaped points, finally if 
$a^{MSSM}_{\mu} \ge 1\times 10^{-9}$ only 52 remain, (the square points).
\begin{figure*}[t!]
\begin{center}
\includegraphics*[scale=0.6]{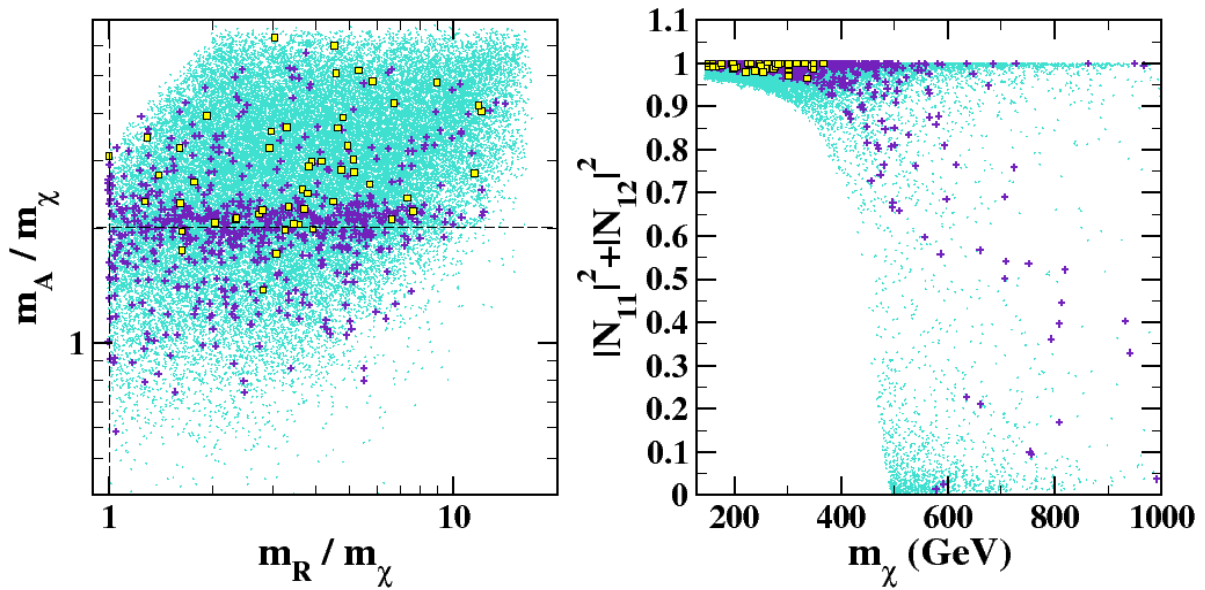}
\caption{
{\it Left:} Scatter plot of the ratio $m_A /m_{\chi}$ versus the
ratio $m_R / m_{\chi}$.  
{\it Right:} Scatter plot of the gaugino fraction 
$|N_{11}|^2 +|N_{12}|^2$ versus the neutralino mass. 
The scanned parameter space and the imposed experimental 
constraints are described in Sec.~\ref{sec1}. The same legend of Fig.~\ref{fig1a} applies. }
\label{fig2a}
\end{center}
\end{figure*}

\section{Neutralino dark matter}
\label{sec2}

The spin-independent neutralino-nucleon cross section 
in the limit of heavy squarks and large $\tan\beta$ 
can be approximated as~\cite{carena1,carena2,carena3,carena4}  
\beq
\sigma^{SI}\simeq \frac{{g'}^2 g^2 |N_{11}|^2 |N_{13}|^2 m_N^4 }{4\pi m_W^2 m_A^4}\tan^2 \beta\times K_{f},
\label{sif}
\eeq
where $N_{11}$ and $N_{13}$ are the lightest neutralino
unitary mixing matrix elements, $m_N$ the nucleon mass (neglecting the mass difference
between the neutron and the proton) and $K_{f} $
a factor which depends  on nucleon form factors.
It scales like $\tan^2 \beta / m_A^4$ and it is 
able to constrain the low $m_A$-high $\tan\beta$ region
even if to a lesser extent than flavor physics observables
that scale like  $\tan^6 \beta / m_A^4$.

The right panel of Fig.~\ref{fig1a} presents the allowed region in the 
($m_A$, $\tan\beta$) plane: the region delimited by the line is excluded by CDF
search in the channel $p\bar{p}\to A+X, A\to \tau^+ \tau^-$.
The left panel of Fig.~\ref{fig1a} shows the scatter plot for the spin-independent 
neutralino-nucleon cross section as a function of $m_\chi$ and the 
region excluded by CDMS~\cite{cdms,cdmsfinal}.
We emphasize that 
CDF and CDMS limits are very mild constraints as can be seen in Fig.~\ref{fig1a}.
The region excluded by CDF is practically excluded by the other constraints
while the CDMS limit exclude only one plus-shaped point (not reported in Fig.~\ref{fig1a})
leaving untouched the regions preferred by WMAP and the $(g-2)_\mu$ anomaly.
Further, the final CDMS upper limits curve exclude around 300 light-gray (magneta) points
between the solid and the dotted line in Fig.~\ref{fig1a} and it is not still 
constraining the more interesting region. For clarity, in all the other plots only the 
final CDMS limits are applied.
Actually, it is more meaningful to compare the CDMS results with the plus-shaped 
and square points: in fact the limit on scattering with nuclei are extracted from rates
which depend on the local density of dark matter 
in our galaxy halo which is assumed to be furnished by the
weakly interacting massive particle, in this case the lightest neutralino. 
From this point of view the negative results of these experiments are
natural in the present scenario and the two events found in the signal region by 
CDMS collaboration cannot be explained by our scenario.

\begin{figure*}[t!]
\begin{center}
\includegraphics*[scale=0.6]{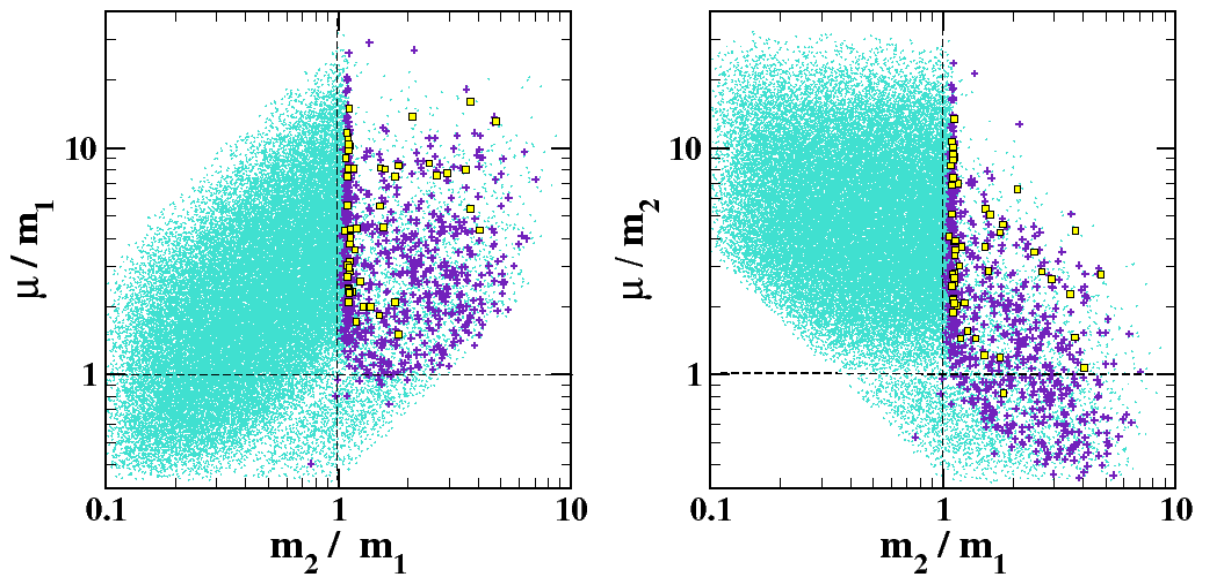}
\caption{
{\it Left:} Scatter plot of the ratio $\mu / m_1$
versus $m_2 / m_1$. 
{\it Right:} Scatter plot of the ratio $\mu / m_2$
versus $m_2 / m_1$. The scanned parameter space and the imposed experimental 
constraints are described in Sec.~\ref{sec1}. The same legend of Fig.~\ref{fig1a} applies. }
\label{fig2b}
\end{center}
\end{figure*}  

The XENON100 experiment~\cite{xenon} should 
reach the sensitivity corresponding to the dashed gray (red) line in the Figure~\ref{fig1a} (left panel).
Such sensitivity is able to cover the region with the highest cross 
section, $m_\chi \ge 300$ GeV, where there is large higgsino component,
as we will discuss below. On the other hand
the region preferred by $(g-2)_\mu$ anomaly cannot be covered.
We also report the prospected sensitivity goal of the XENON experiment 
with 1 ton detector mass~\cite{xenon}, dot-dashed grey (red) line, which is 
$10^{-11}-10^{-10}$ pb for neutralino mass in the range 
$100-1000$ GeV: practically all of the parameter space 
will be probed. 

As no relation has been imposed between the neutralino mass and $m_A$
and between gaugino mass parameters,
it is interesting  to explore which correlations may emerge by the imposition of 
all the applied constraints.

Fig.~\ref{fig2a}, left panel, presents the scatter plot 
of the ratio $m_A /m_{\chi}$ versus $m_R / m_{\chi}$
where $m_R$ is right-right mass parameter for the stau.
Points with the correct relic density abundance 
accumulate along the line $m_A /m_{\chi} \simeq 2$ 
where 
neutralino pair annihilation into fermions through resonant
$s$-channel exchange of neutral Higgs bosons $A$, $H$ 
is the dominant mechanism in large portion of the parameter space.
Stau coannihilation is at work for models where $m_R \sim m_\chi $ and coannihilation 
with the second neutralino and the lightest chargino are important for larger values 
of the ratios.

In the right panel of Fig.~\ref{fig2a} the
gaugino fraction $|N_{11}|^2 +|N_{12}|^2$ is plotted against the
neutralino mass. For neutralino masses below 400 GeV, the preferred region
by the $(g-2)_\mu$,  is pure gaugino
while for masses greater than 400 GeV higgsino component is present.
This effect can be seen in Fig.~\ref{fig1a} in the spin-independent 
neutralino cross section which depends on $N_{13}$ 
($|N_{13}|^2 +|N_{14}|^2 = 1-|N_{11}|^2 +|N_{12}|^2$): 
the models with the highest cross section 
are the ones with $m_{\chi} \ge 400$ GeV where the $\chi\chi \Phi$
coupling is enhanced in reason of a  larger higgsino component.

The left and right panels of Fig.~\ref{fig2b} present the scatter plot in the 
($\mu/m_1$, $m_2/m_1$) and ($\mu/m_2$, $m_2/m_1$) planes respectively.
We see that models with the correct relic abundance have  $m_2 \ge m_1$ 
$\mu \ge m_1$. The $(g-2)_\mu$ prefer models with $\mu \ge m_2$.        
The models with strong gaugino-higgsino mixing, $m_1 \simeq \mu$, $\mu \le m_2$,
$|N_{11}|^2 +|N_{12}|^2 \le 0.9$ can be probed by XENON100. 
We further note that most of the point in WMAP and $(g-2)_\mu$ ranges
are charactherized by having high degeneracy in the gaugino masses $m_1 \simeq m_2$.
Such conditions give a ``well-tempered bino/wino'' neutralino which can be realized 
in model with split supersymmetry as shown for example in Ref.~\cite{arkani}

\section{Prospects for LFV signals}
\label{sec3}

\begin{figure*}[t!]
\begin{center}
\includegraphics*[scale=0.6]{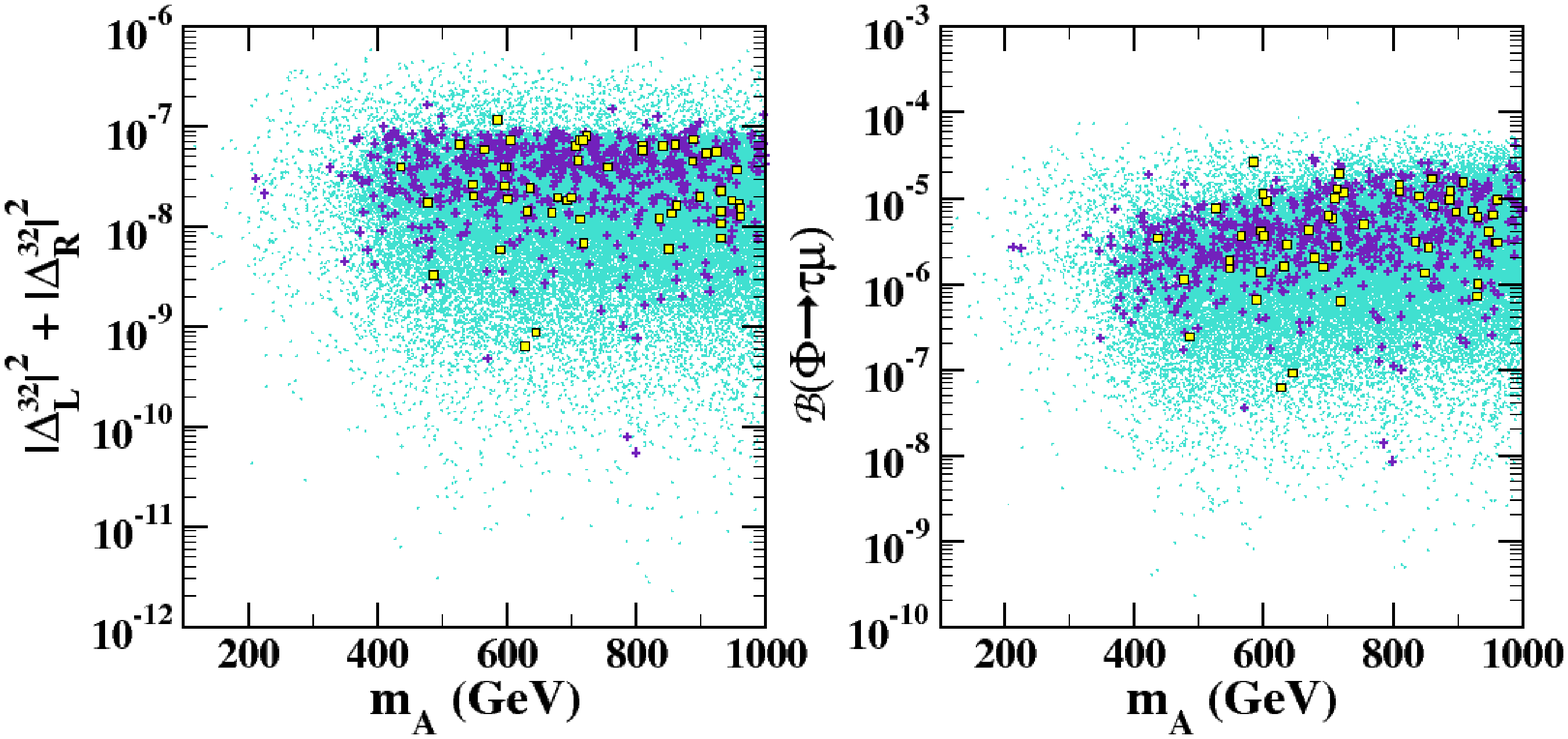}
\caption{{\it Left:} Scatter plot of the LFV violating vertex in Eq.~(\ref{effv}) as a function of $m_A$. 
{\it Right:} Scatter plot of the branching ratio $\Phi\to\tau\mu$ versus $m_A$ ($\Phi = A,\,H$). 
The scanned parameter space and the imposed experimental 
constraints are described in Sec.~\ref{sec1}. The same legend of Fig.~\ref{fig1a} applies.}
\end{center}
\label{fig3a}
\end{figure*}

The $\tau$ LFV $\tau$ decay which is more sensible
to Higgs mediated effects is $\tau\to\mu\eta$~\cite{sher,brignole2,kanemura2,paradisi2,
chuan,arganda2,herrero} and the branching ratio reads~\cite{brignole2,paradisi2}:
\beq
\frac{{\cal B}(\tau \to \mu \eta)}{{\cal B}(\tau \to \mu \nu \bar{\nu})}
 = 9\pi^2 \frac{f_{\eta}^2 m_{\eta}^4}{m_{\tau}^2 m_A^4}F_{\eta}^2
\left(1-\frac{m_{\eta}^2}{m_{\tau}^2}\right)^2  \Delta^2 \tan^6 \beta. 
\label{btme}
\eeq
Here $f_\eta \simeq 110$ MeV, 
${\cal B}(\tau \to \mu \nu \bar{\nu}) =\frac{G_F^2 m_{\tau}^5}{192 \pi^3}\frac{1}{\Gamma}$
and $F_\eta$ a factor which depends on the hadronisation of quarks bilinears matrix elements: in the treatment
of Ref.~\cite{brignole2} it is such that $F_\eta^2 \sim 2$.\footnote{The approach
using chiral perturbation theory~\cite{arganda,herrero} gives results
different at most by a factor two. These uncertainties will not change our conclusions.} 
Lepton flavor violation  enters through the factor
\beq 
\Delta^2 =|\Delta_L^{32}|^2 + |\Delta_R^{32}|^2.
\label{effv}
\eeq
We also consider the radiative decay $\tau \to \mu\gamma$ which receives
also important contributions by gaugino-slepton loop diagrams:
the factors $|\Delta_L^{32}|^2$
and $|\Delta_R^{32}|^2$ enter separately in the branching ratio and not through the
combination in Eq.~(\ref{effv}). For the computation we used the formulas in Ref.~\cite{paradisi1}
including both gaugino mediated and Higgs mediated effects.

\begin{figure*}[t!]
\begin{center}
\includegraphics*[scale=0.6]{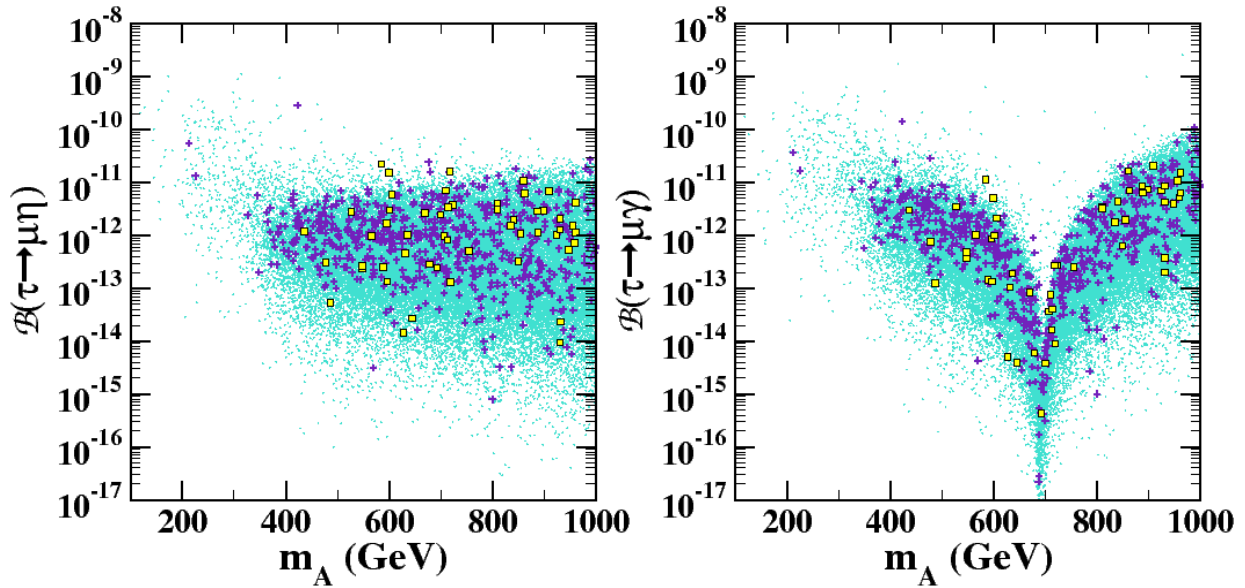}
\caption{{\it Left:} Scatter plot of the LFV violating vertex in Eq.~(\ref{effv}) as a 
function of $m_A$. 
{\it Right:} Scatter plot of the branching ratio $\Phi\to\tau\mu$ versus $m_A$ ($\Phi = A,\,H$). 
The scanned parameter space and the imposed experimental 
constraints are described in Sec.~\ref{sec1}. The same legend of Fig.~\ref{fig1a} applies.}
\label{fig3b}
\end{center}
\end{figure*}

The CP-odd Higgs boson decay width and  branching ratio
are~\cite{cannoni}
\beqa
\Gamma({A\to\tau^+\mu^-})&=&
\frac{1}{2} \tan^{2}{\beta} \, \Delta^2 \,  \Gamma({A\to \tau^+\tau^-}),\cr
{\cal B}(A\to \mu^+ \tau^-)&+&{\cal B}(A\to \mu^- \tau^+)\cr
&=& \tan^{2}{\beta} \, \Delta^2 \, {\cal B}(A\to \tau^+ \tau^-),
\label{hdecay}
\eeqa
where $\Delta^2$ is defined in Eq.~(\ref{effv})
and we used the fact that $\Gamma({A\to \tau^+\mu^-})=\Gamma({A\to \tau^-\mu^+})$.
For the CP-even Higgs boson $H$, the right hand sides of Eq.~(\ref{hdecay}),
are multiplied by a factor ${\sin}(\beta - \alpha)/(\cos\alpha)^2$ which is order one in our 
scenario where $m_A \simeq m_H $, thus the previous formulas hold for both bosons. 

In Fig.~\ref{fig3a}, left panel, we present the scatter plot for the effective vertex given by 
Eq.~(\ref{effv}),
while the right panel shows the scatter plot of the branching ratio
given by Eq.~(\ref{hdecay}) as a function of $m_A$.
We see that they reach $10^{-7}$ and $3\times 10^{-5}$ respectively for 
models preferred by WMAP measurements, two orders of magnitude lower than 
what found without imposing it on the parameter space~\cite{brignole1,moretti,cannoni}. 
The branching ratios of LFV decays $\tau\to\mu\eta$ and 
$\tau\to\mu\gamma$ versus $m_A$ are given in Fig.~\ref{fig3b}.
For models with the correct neutralino relic density abundance and preferred
by the $(g-2)_\mu$ anomaly, both are generally under $10^{-10}$ while relaxing 
the constraints lower bound they can reach the $10^{-9}$ level. 
We remind that 
at a Super-B 
factory the present limits ${\cal O}$($10^{-8}$) can be lowered
to ${\cal O}$($10^{-9}\sim 10^{-10}$) for the $\mu\eta$ final state because 
the branching ratio scales linearly with the luminosity due to practically
negligible background. In the $\mu\gamma$ case for the presence of large
background the branching ratio scale as the square root of the luminosity
and the sensitivity reach is one order of magnitude lower~\cite{belle}.

We further revisit the prospects for detection of Higgs
LFV signals in $pp$ collisions at LHC~\cite{brignole1,moretti} 
and in $\gaga$ collisions at the photon collider option
of the future International Linear Collider~\cite{cannoni}. 

At high $\tan\beta$ the dominant production mechanisms
for $A,H$ at LHC is $b\bar{b}$ fusion
due to the $m_b \tan\beta$ enhanced $b\bar{b}\Phi$ couplings.
We calculate the cross section with \textsc{FeynHiggs} 
which uses the approximation  
\beq
\sigma^{MSSM}(b\bar{b}\to \Phi) = \sigma^{SM} (b\bar{b}\to \Phi ) 
\frac{\Gamma(\Phi \to b\bar{b})^{MSSM}}{\Gamma(\Phi \to b\bar{b})^{SM}},
\eeq
where $\sigma^{SM}(b\bar{b}\to \Phi)$ is the total SM cross section
for production of Higgs boson with mass $m_{\Phi}$ via $b\bar{b}$ fusion: to obtain the 
value in the 
MSSM it is rescaled with the ratio of the decay width of the inverse process in
the MSSM over the SM decay width~\cite{feynhiggs,hein1,hein2}.
We calculate for each random model the product the
$\sigma(pp\to \Phi +X)\times{\cal B}(\Phi\to\tau\mu)$.
As masses and couplings of $A$ and $H$ 
are practically identical as discussed above, we have 
$\sigma(pp\to A+X) + \sigma(pp\to H+X)\simeq 2 \sigma(pp\to A+X)$.

\begin{figure*}[t!]
\begin{center}
\includegraphics*[scale=0.6]{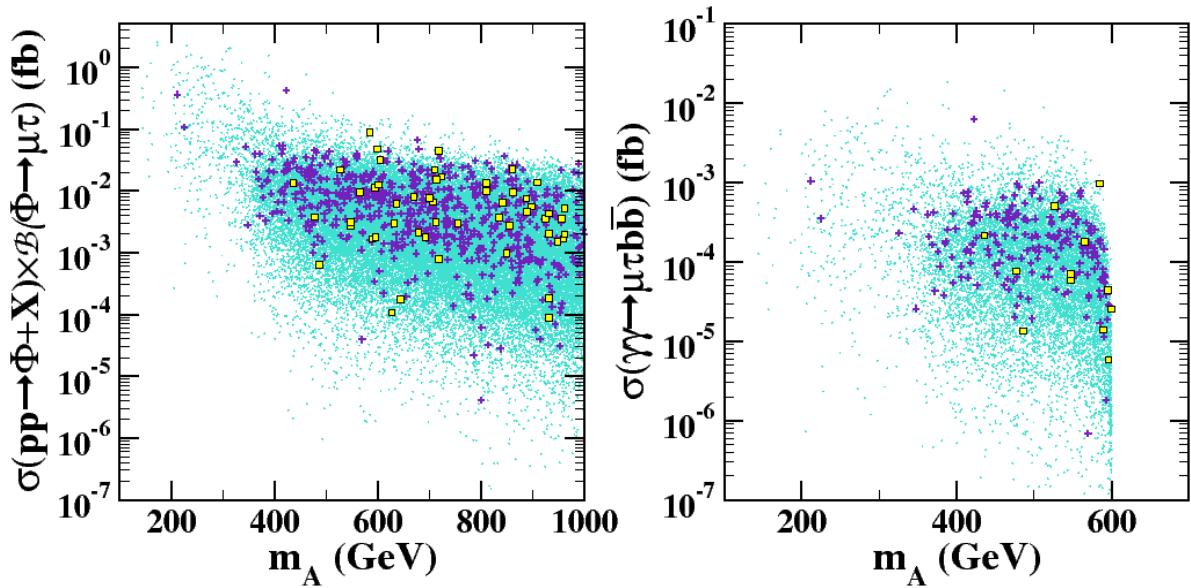}
\caption{
{\it Left:} Scatter plot of the inclusive production cross  
section $pp\to\Phi +X$ times the branching ratio of $\Phi\to\tau\mu$ at LHC 
versus $m_A$ ($\Phi =A, \,H$). 
{\it Right:} Scatter plot for the cross section of the process $\gaga\to\tau\mu b\bar{b}$
in photon-photon collision at $\sqrt{s_{\gaga}} = 600$ GeV. 
The scanned parameter space and the imposed experimental 
constraints are described in Sec.~\ref{sec1}. The same legend of Fig.~\ref{fig1a} applies.}
\label{fig4a}
\end{center}
\end{figure*}

The scatter plot 
$\sigma(pp\to \Phi +X)\times{\cal B}(\Phi\to\tau\mu)$ is shown in Fig.~\ref{fig4a}, left panel. 
We see that with 
the nominal integrated luminosity of $100$ fb$^{-1}$ per year models 
which satisfy both the relic density abundance and $\Delta a_\mu$ 
can give up to 10 events per year (squared points), up to 40 if we relax the condition
on the lower limit of $\Delta a_\mu$ (plus-shaped points) and up to 200-300 relaxing
both the lower limits (magenta (grey)  points).

In $\gaga$ collisions the main production mechanism for $\Phi =A,H$
is $\tau\tau$ fusion~\cite{choi} while the $b\bar{b}$ is suppressed
by a factor $3(1/3)^4 (m_b /m_{\tau})^2 \simeq 0.1$ which cannot
be compensated by corrections to the $b$ Yukawa coupling.
In Ref.~\cite{cannoni} we studied in detail
the $\mu\tau$ fusion process $\gamma\gamma \to \mu\tau b\bar{b}$
where the Higgs boson is produced in the $s$-channel via a virtual $\mu\tau$ pair and
can be detected from its decay mode $A \to b\bar{b}$.  

A good analytical approximation for the cross section is obtained using
the equivalent particle approximation wherein the colliding
real photons split respectively into $\tau$ and $\mu$ pairs with the subsequent
$\mu\tau$ fusion into the Higgs boson. The splitting functions
of the photon at leading order read~\cite{choi}:
\beq
P_{\gamma/\ell}(x)=\frac{\alpha}{2\pi}[x^2 +(1-x)^2]
\ln\left(\frac{m_\Phi ^2}{m_\ell^2}\right),
\eeq
thus the cross section is given by:
\beqa
&&\sigma(\gaga \to \mu\tau b\bar{b}; s_{\gaga})= \frac{4
\pi^2}{s_{\gaga}}
\frac{\Gamma(A\to \tau\mu) {\cal B}(A\to b\bar{b})}{M_A}\cr
&&\times 2\int_{-\ln 1/t}^{+\ln 1/t} {d\eta}\; P_{\gamma/ \mu}\left(t e^\eta \right)
P_{\gamma /\tau}\left(t e^{-\eta}\right),
\label{fusionfinal}
\eeqa
with $t={m_A}/{2E_\gamma} $, $\eta = \ln\sqrt{{x_\mu}/{x_\tau}}$,
$x$ is the fraction of the energy of the photon carried by the
virtual lepton. 
In~\cite{cannoni} we have shown that the effect of photons spectra 
can be neglected, we thus consider  
monochromatic photons with $\sqrt{s_{\gaga}} =600 $ GeV,
and photon-photon luminosity 
500 fb$^{-1}$ yr$^{-1}$
based on the parameters of
TESLA(800)~\cite{tesla}.

The scatter plot of the signal cross section versus $m_A$ is shown in Fig.~\ref{fig4a},
right panel. Here the models 
which satisfy both the relic density abundance and $\Delta a_\mu$ (squared points)
have maximal cross section $10^{-3}$ fb, which is too small.
Relaxing the lower limits cross section values up to $2\times 10^{-2}$ fb are possible,
giving 10 events/year.

\section{Summary and conclusions}
\label{sec4}

In the framework of the MSSM with heavy SUSY-QCD
particles and large $\tan\beta$ 
we have studied lepton flavor violation mediated by
the heavy neutral Higgs $\Phi=A$, $H$
in $\tau-\mu$ sector both in $\tau$ lepton decays,
$\tau\to\mu\eta$ and $\tau\to\mu\gamma$, and at high energy collider through the production 
and decay at LHC, $p p \to \Phi +X$, $\Phi \to \tau\mu$ and the $\mu-\tau$ fusion at 
a photon collider, $\gaga\to\tau\mu b\bar{b}$. The approach to LFV has been model independent 
by the use of the mass insertion approximation. We used large mass insertions 
$\delta^{ij}_{LL} = \delta^{ij}_{RR} =0.5$ to estimate the number of events
in the most favourable scenario that can be obtained in future experiments. 
With such a large source of LFV  the branching ratios
of rare decays can exceed the present experimental upper bounds from BABAR and BELLE which 
therefore provide 
constraints on the MSSM parameter space in presence of LFV. Other constraints that have been imposed 
are limits from direct search of sparticles and of the light Higgs, $B$-physics observables, 
the $(g-2)_{\mu}$ anomaly, and recent limits from TEVATRON search of non standard Higgs bosons
in the $\tau\tau$ channel. In the R-parity conserving MSSM the heavy neutral Higgs
play an important role both in neutralino annihilation cross sections to satisfy the
relic density of dark matter measured by WMAP and in the spin-independent neutralino nucleon
cross section in direct dark matter search experiment. We thus have imposed on the MSSM parameter space
the present CDMS      
exclusion limit and the WMAP limits on $\Omega h^2$.

We have found that in models with $0.09 \le \Omega_{\chi} h^2 \le 0.13$ 
and $1\le a_{\mu}^{MSSM}\times 10^{9} \le 4$: the branching ratios of $\tau\to\mu\eta$, 
$\tau\to\mu\gamma$ are both under the ${\cal O}(10^{-10})$ thus probably 
undetectable even at super-B factory; at LHC the cross section for   
$p p \to \Phi +X$, $\Phi \to \tau\mu$ can reach ${\cal O}(10^{-1}-10^{-2})$ fb in the
range $m_A =400-1000$ GeV giving up to 10 events with 100 fb$^{-1}$; 
the cross section of $\gaga\to\tau\mu b\bar{b}$ reach ${\cal O}(10^{-3})$ fb,
thus too small even for the large value of the expected luminosity of 500 fb$^{-1}$.
Prospects are somewhat more encouraging if we relax the lower limits, imposing only
$\Omega_{\chi} h^2 \le 0.13$ and $a_{\mu}^{MSSM}\times 10^{9} \le 4$. In this case 
branching ratios of LFV $\tau$ decays can reach ${\cal O}(10^{9})$, 
the cross sections at LHC about 2 fb for low $m_A$ masses and around $2\times 10^{-2}$ fb
in $\gaga$ collisions. 

We derive two main indications from this analysis. On one hand, even with large sources of lepton
flavor violation in the slepton mass matrix, the process where Higgs mediated
$\tau-\mu$ transitions should manifest could be beyond the sensitivity reach of future experiments.
On the other hand, to observe such effects, in any case, the full luminosity of the machine is needed.


We emphasize that our not optimistic conclusions are specific to Higgs mediated effects.
As shown in~\cite{barger1}, in typical SUSY parameter space where gaugino-mediated effects 
are dominant over Higgs mediated ones and in the context of SUSY see-saw mechanism,
LFV rates are detectable by future experiments.


We have also studied the spin-independent neutralino nucleus cross section: we have shown that   
in models that satisfy  $0.09 \le \Omega_{\chi} h^2 \le 0.13$ and $1\le a_{\mu}^{MSSM}\times 10^{9} \le 4$,
the cross section lies just below the sensitivity of XENON100 which should report 
results soon. The full XENON 1 ton is needed to cover all the parameter
space. However, if the lower  limit  on $(g-2)_{\mu}$ is not considered XENON100 
is sensitive to the neutralino mass range 300-1000 GeV in models where the higgsino 
component is large.

\end{document}